\documentclass[10pt,twocolumn,pra,showpacs,floatfix]{revtex4}
\usepackage{graphicx}
\usepackage{subfigure}
\usepackage{bm}
\begin{document}

\title{The role of light ellipticity in ionization of atoms by intense few-cycles laser pulses }

\author{M. Abu-samha and L.~B. Madsen}
\affiliation{
Lundbeck Foundation Theoretical Center for Quantum System Research, Department of Physics and Astronomy,
Aarhus University, DK-8000 Aarhus C, Denmark.}
\begin{abstract}
We provide theoretical investigations of the response of the Ar and H atoms to an intense elliptically polarized few-cycle
laser pulse, as a function of light ellipticity. The time-dependent Schr{\"o}dinger equation describing the least-bound
electron is solved numerically, and differential quantities such as the momentum distribution,
the electron density in the continuum, and the above-threshold ionization spectra are computed. These quantities provide
insight into the ionization dynamics and the electron rescattering process as a function of light ellipticity, and reveal great similarities
between the response of Ar and H to the applied external field.

\end{abstract}

\maketitle
\section{Introduction}
Experiments are now emerging on strong-field ionization of atoms~\cite{P.Eckle12052008} and molecules~\cite{staudte:033004,H.Akagi09112009,Lotte2010}
by femtosecond, elliptically polarized lasers. For an elliptically polarized laser, the field has two perpendicular components,
and the electron dynamics and the resultant momentum distributions depend on the relative magnitudes of the field components
(light ellipticity), pulse duration (few-cycle vs. many-cycle ionization regimes), and carrier-envelope phase (CEP), see,
for example, Refs.~\cite{martiny:043404,martiny:043416,martiny:093001}
and references therein.

In a recent experiment~\cite{P.Eckle12052008} on strong-field ionization of the He atom
by a few-cycle elliptically polarized laser, the momentum
distributions in the polarization plane of the external field are rotated slightly
relative to the prediction of the simple-man's model, in which the final momentum of a continuum
electron born at time $t$ is defined (in atomic units, which are used throughout) as $\vec{k}_{f}=-\int_{t}^{\infty} \vec{E}(t^\prime)dt^\prime=-\vec{A}(t)$
where $\vec{E}(t)$ is the electric field and $\vec{A}(t)$ the vector potential at time $t$ (cf. Eq.~\ref{Eq1}). For a few-cycle pulse of duration $T$, electron
emission at the peak of the pulse ($T/2$) dominates the ionization process, and the final momentum is thus  $\vec{k}_{f}=-\vec{A}(T/2)$.
In~\cite{0953-4075-42-16-161001}, calculations of
the response of the H atom to a few-cycle circularly polarized laser pulse, based on solution
of the time-dependent Schr{\"o}dinger equation (TDSE), produced a similar angular shift.
The angular shift has been attributed to the interplay between the Coulombic potential and
the external field~\cite{0953-4075-42-16-161001}.

In the present work, we investigate the response of many-electron atoms
to intense few-cycle elliptically polarized laser pulses. We consider Ar
because its ionization potential is comparable to that for the H atom, and
this would facilitate comparisons with the recent TDSE results
for the latter~\cite{0953-4075-42-16-161001}. Note that
H and Ar produce very similar momentum distributions when probed by
a linearly polarized few-cycle laser pulse~\cite{0953-4075-41-24-245601},
because the ionization process is not sensitive to the short-range potential (the precise electronic structure of the probed atom).
Here, we extend the investigation to elliptically polarized fields, with light ellipticity [$\epsilon$ in Eq.~(\ref{Eq1})] ranging from
$\pi$/2 (circular polarization) to $\pi$/10 (approaching the linear polarization limit). Note that whereas circular polarization switches off electron rescattering,
electron rescattering effects are very important in the linear polarization limit~\cite{PhysRevLett.71.1994}.
Here, we discuss the effect of light ellipticity on the momentum distributions.
Our analysis is assisted by above-threshold ionization (ATI) spectra and electron density in the continuum.
These quantities help us determine at which $\epsilon$ value electron rescattering becomes important, and this has implications
on the application of elliptically polarized lasers to investigations of orbital structure and symmetry~\cite{Martiny2010,Lotte2010}.

The following are the main findings of this work. The momentum distributions
of Ar and H show great similarities between the responses of the two systems
to the external field, and to change of light ellipticity. For both atoms, the momentum
distributions are rotated relative to the predictions of the simple-man's model. The angular shift does not
depend very much on the probed system, indicating that in the present regime it is determined mainly
by the long-range part of the atomic potential. The angular shift is very sensitive to light ellipticity, and it is smallest for the circularly polarized light.
By investigating the electron density in the continuum
and the ATI spectra as a function of light ellipticity, we find that electron rescattering becomes important at $\epsilon\le\pi$/4.

The paper is organized as follows. The computational details are given in Sec.~\ref{CompDet}, the
results and discussion in Sec.~\ref{Res}, and summary and conclusions in Sec.~\ref{Conc}.

\section{Computational Details}
\label{CompDet}
The external field is defined as $\vec{E}(t)=-\partial_t\vec{A}(t)$,
where $\vec{A}(t)$ is the vector potential [defined in the $xy-$plane,
cf. Eq.~(\ref{Eq1})]. For an elliptically polarized light, we write the
vector potential as
\begin{equation}
\label{Eq1}
 \vec{A}(t)=A_0f(t)\left(  \begin{array}{c}\cos(\omega t + \phi)\cos(\epsilon/2)
                                        \\ \sin(\omega t + \phi)\sin(\epsilon/2) \\ 0 \end{array}    \right),
\end{equation}
where $A_0$ is the amplitude, $\omega$ the carrier frequency, $\phi$ the CEP value, $\epsilon$ the light ellipticity,
and $f(t)=\sin^2 (\omega t/2N)$ the envelope for an $N$-cycle pulse. In the present study, $\omega=0.057$~a.u. (800 nm
wavelength), $\phi=-\pi/2$, and $N=3$ optical cycles.

The wave function is expressed in spherical harmonics as
\begin{equation}
\label{Eq2}
\Psi({\bm r}, t)=\sum_{l=0}^{l_{max}}\sum_{m=-l}^{l} \frac{f_{lm} (r)}{r} Y_{lm}(\Omega),
\end{equation}
and the TDSE is solved in the velocity gauge with a grid representation for the reduced radial
wave functions $f_{lm}(r)$~\cite{kjeldsenPRA07}.
The SAE potential describing Ar is taken from Ref.~\cite{MullerPRL98}.
We use an equidistant grid with 4096 points that extends up to 400~a.u. for Ar (300~a.u. for H).
Truncating the expansion in (\ref{Eq2}) at $l_{max}=$ 40 for Ar (35 for H) produces
converged results at laser intensity of 1.06$\times$10$^{14}$~W/cm$^{2}$. We note that for
each $l$, the azimuthal quantum number runs through 2$l$+1 values making the calculations
for an elliptically polarized light fully 3D and much more time demanding than in the case
of linearly polarized light.

The calculations are carried out at $\epsilon=$ $\pi$/2, $\pi$/3, $\pi$/4, and $\pi$/10.
 The momentum distributions are computed in the polarization plane ($dP/d\vec{k}$ with $\theta_k$ fixed at $\pi/2$) by projecting on
scattering states~\cite{madsen:063407}. Since the external field (\ref{Eq1}) is defined in
the $xy-$plane, we only consider contributions from the Ar 3p$_x$ and 3p$_y$
states. For symmetry reasons, the 3p$_z$ orbital will not contribute to ionization in the $xy-$plane.

\section{RESULTS AND DISCUSSION}
\label{Res}

\begin{figure*}
{\includegraphics[width=0.9\textwidth]{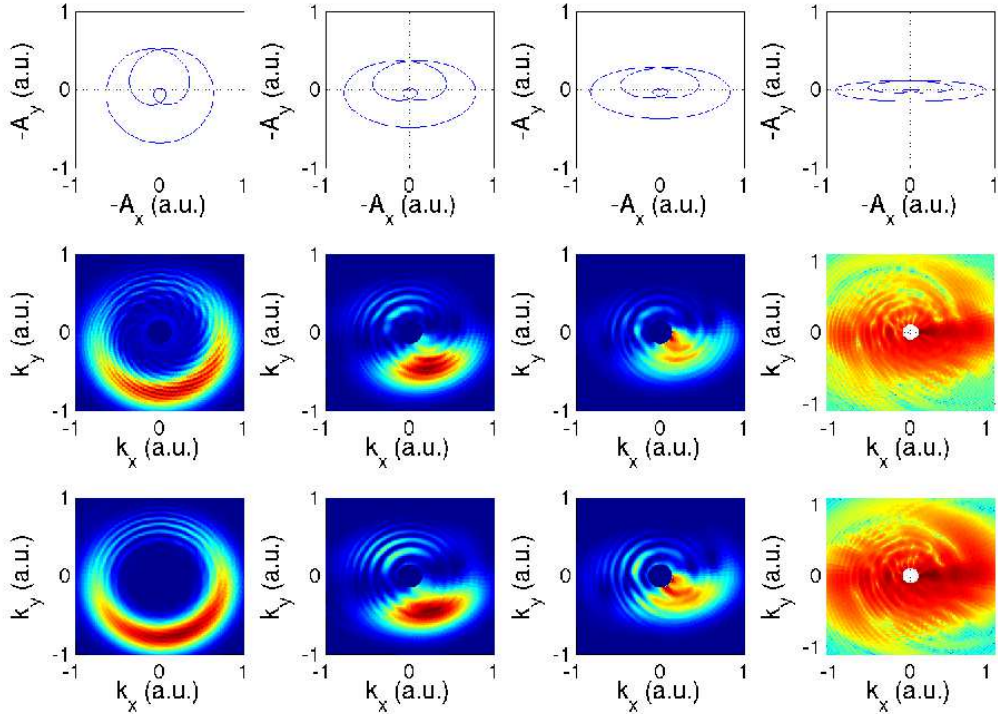}}
\caption{(Color online) 2D momentum distribution for Ar (middle panels) and H (bottom panels) in an elliptically
polarized field (vector potential in top panels) with $\epsilon=\pi/2$, $\pi/3$, $\pi/4$, and $\pi/10$ from left to right, respectively.
The laser pulse contains 3 cycles and has a peak intensity 1.06$\times$10$^{14}$W/cm$^{2}$ and frequency 0.057~a.u.
In the rightmost panels, the $dP/d\vec{k}$ scale is logarithmic. }
\label{fig1}
\end{figure*}

In Fig.~\ref{fig1}, we provide parametric plots of the vector potentials (top panels) and momentum distributions
in the $xy-$plane ($dP/d{\vec k}$ at $k_z=0$ ) for Ar (middle panels) and H (bottom panels) in elliptically polarized fields with fixed intensity
of 1.06$\times$10$^{14}$~W/cm$^{2}$ and light ellipticity $\epsilon=\pi/2$, $\pi/3$, $\pi/4$, and $\pi/10$,
from left to right, respectively. We set $dP/d{\vec k}=0$ for $k_x^2$+$k_y^2 \le 0.01$~a.u., in order
to achieve a better graphical display. The momentum distributions for Ar are incoherent sums of the individual
contributions from the 3p$_x$ and 3p$_y$ states.
At $\epsilon=\pi/2$, $\pi/3$, and $\pi/4$, both states contribute to the momentum distributions.
At $\epsilon=\pi/10$, by contrast, the major polarization axis of the external field is aligned with
the $x-$axis as can be seen in Fig.~\ref{fig1}, and the contribution from the 3p$_x$ state
dominates the momentum distribution.

Starting with Ar at $\epsilon=\pi/2$, the momentum distribution shows a ring-like structure
with large emission probability in a finite momentum region [$k_y\approx 0.75$~a.u.], characteristics of a short pulse
duration~\cite{martiny:043416}. The momentum distribution shows a spiral structure that develops
outward with a counter clockwise rotation, following the time evolution of the vector potential (\ref{Eq1}).
Moreover, the momentum distribution is characterized by an angular shift, relative
to the predictions of the final momentum based on the simple-man's model ($\vec{k}_{f}=-\vec{A}(T/2)$). As mentioned earlier, this shift is due to
the interplay between the atomic potential and the external field~\cite{0953-4075-42-16-161001}.
From the momentum distribution, one can see that the probability of finding electrons with low momenta (0.1 $< \vec{k}_{f} < $  0.5~a.u.) is
very small, indicating that the continuum electron is progressively driven away from the core by the external
field, and that the electron rescattering channel is closed.

In fact, it has recently been shown by both experiment and calculations that investigations of oriented samples
(atomic~\cite{Martiny2010} and molecular~\cite{Lotte2010}) by a circularly polarized light (for which electron rescattering is negligible) provide a
unique probe of the orbital symmetry$-$the orbital angular nodes are preserved in the momentum distributions.

For Ar at $\epsilon=\pi/3$, the momentum distribution is still characterized by a
large emission probability in a finite momentum region. However, at a progressively lower momentum compared
to the calculations at $\epsilon=\pi/2$. This latter feature is understood based on the simple-man's prediction
of the final momentum. One can also see that the probability of finding low-momentum
electron is still small, compared to the main emission probability. We note that the angular shift of
the momentum distribution increases by going to smaller $\epsilon$ values.
Our TDSE results show that the angular shifts for Ar are $10^\circ$ at $\epsilon=\pi/2$ and $27^\circ$
at $\epsilon=\pi/3$.

Turning to the $\epsilon=\pi/4$ case, the regular emission pattern observed at $\epsilon=\pi/2$ and $\pi/3$ is now distorted
by rescattered electrons and bears a clear signature of low-energy electrons typical for tunneling
ionization by the linear component of the external field. Because of these new features, it is no longer meaningful to
discuss the angular shift at $\epsilon=\pi/4$. At $\epsilon=\pi/10$, the momentum distribution is mainly along the $k_x-$axis. However,
it is not inversion symmetric across $k_y$. Moreover, the radial jets and the interference structure typical for ionization
by a linearly polarized laser pulse~\cite{0953-4075-41-24-245601} are washed out by the circular component of the external field.

The momentum distributions computed for H (bottom panels of Fig.~\ref{fig1}) are generally similar to those for Ar, and show the same
response to change of light ellipticity. For instance, the angular shifts computed for H  ($9^\circ$ at $\epsilon=\pi/2$ and $27^\circ$
at $\epsilon=\pi/3$) are similar to those for Ar, indicating that the shift is not sensitive to the short-range potential of the 
probed atom. There are, nevertheless, quantitative differences between Ar and H 
such as the lack of spiral structure in the momentum distribution of H at $\epsilon=\pi/2$. This is
due to the different ionization potentials (0.5~a.u. for H; 0.57~a.u. for Ar). For a circularly polarized light with
laser intensity of $1.06\times 10^{14}$~W/cm$^{2}$, the Keldysh parameter~\cite{Keldysh} value
for Ar (H) is 1.6 (1.5). The spiral structure has been reported for H at laser intensity $5\times 10^{13}$~W/cm$^{2}$~\cite{0953-4075-42-16-161001},
corresponding to a Keldysh parameter value of 2.1.

\begin{figure*}
{\includegraphics[width=0.9\textwidth]{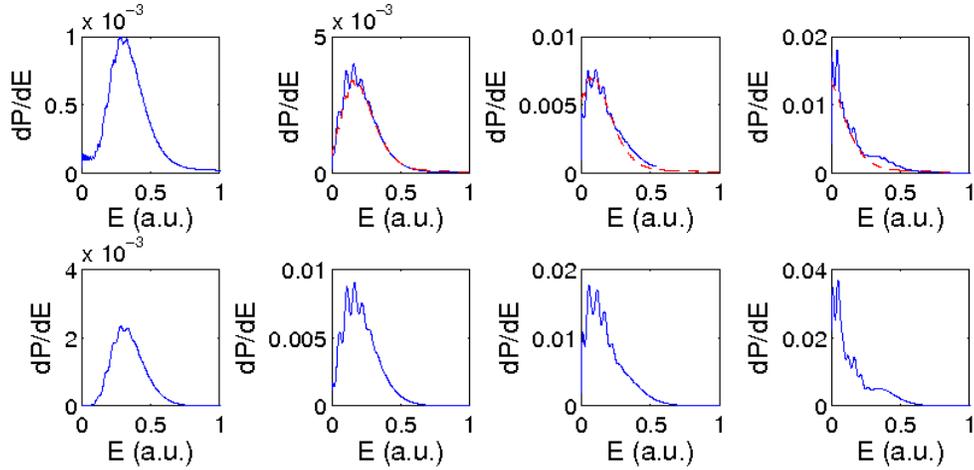}}
\caption{ATI spectra (dP/dE in a.u.) for Ar (top panels) and H (bottom panels) in an elliptically polarized field with
$\epsilon=\pi/2$, $\pi/3$, $\pi/4$, and $\pi/10$ from left to right, respectively. The dashed curves in the top panels denote the ATI spectrum computed
at $\epsilon=\pi/2$, shifted in energy and renormalized. See caption of Fig.~\ref{fig1} for details regarding laser parameters.}
\label{fig2}
\end{figure*}

The ATI spectra, obtained by integrating the 3D momentum distributions
over the angular variables, are shown for Ar and H in Fig.~\ref{fig2}, and provide further support
for our interpretations of the momentum distributions in Fig.~\ref{fig1}. For instance,
for Ar and H at $\epsilon=\pi/2$ (and still at $\pi/3$), the ATI spectra show essentially a single-peak distribution, in agreement
with the predictions of the ionization models~\cite{Reiss80,Reiss87,Popov04}. At $\epsilon=\pi/4$, by contrast, the ATI spectra show signature
of low-energy electrons, characteristics of tunneling ionization by the linear component of the external field. By comparing with the ATI
spectrum obtained at $\epsilon=\pi/2$ and focusing on the high-energy tail, we find that while electron rescattering effects are minimal at $\epsilon=\pi/3$,
they enhance the high energy tail of the ATI spectrum at $\epsilon=\pi/4$. At $\epsilon=\pi/10$, the ATI peak develops a high-energy shoulder due to rescattered electrons.

\begin{figure*}
{\includegraphics[width=0.9\textwidth]{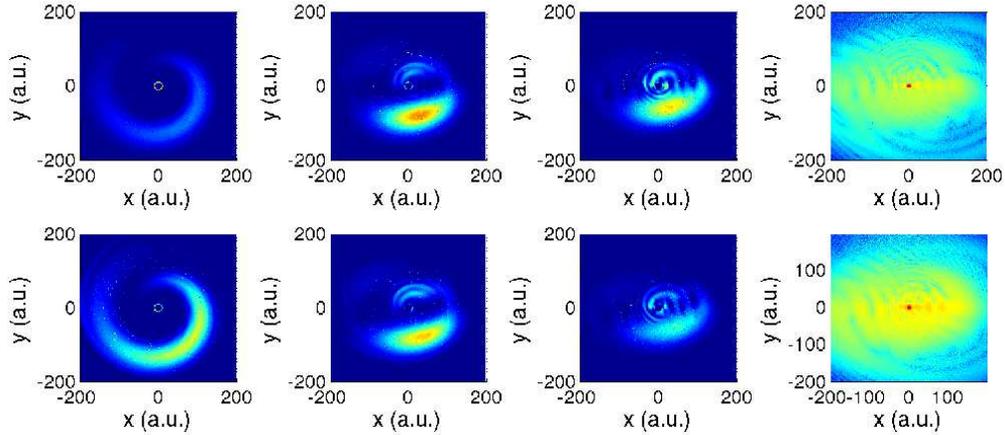}}
\caption{(Color online) Electron density in the $xy-$plane for Ar (top panels) and H (bottom panels) just after the
end of an elliptically polarized pulse with $\epsilon=\pi/2$, $\pi/3$, $\pi/4$, and $\pi/10$ from left to right, respectively.
See caption of Fig.~\ref{fig1} for details regarding laser parameters. In the rightmost panels, the density scale is logarithmic.}
\label{fig3}
\end{figure*}

The electron density in the continuum is shown for Ar and H in Fig.~\ref{fig3}, just after the end
of the laser pulse. At $\epsilon=\pi$/2 and $\pi$/3, the electron density of the continuum electron
is driven away from the core. By contrast, at $\epsilon=\pi$/4, the electron density is smeared out due
to rescattering. At $\epsilon=\pi$/10 the electron density is focused along the major polarization
axis ($x-$axis) of the external field. 

From the analysis of the ATI spectra (Fig.~\ref{fig2}) and electron density in the
continuum (Fig.~\ref{fig3}), we learn that electron rescattering is essentially
absent for elliptically polarized laser pulses with $\pi/3 \le \epsilon \le \pi/2$.
These findings are important for guiding future experiments on the investigation
of orbital structure by employing elliptically polarized laser pulses,
which would benefit from a clean orbital signal$-$without the contribution from
rescattered electrons.

\section{Summary and conclusions}
\label{Conc}
To summarize, we investigated ionization of atomic targets (H and Ar)
by intense 3-cycle elliptically polarized laser pulses with light ellipticity ($\epsilon$)
ranging from the circular polarization case ($\epsilon=\pi$/2) and approaching the
linear polarization case ($\epsilon=\pi$/10). The momentum distributions, ATI spectra,
and electron density in the continuum are studied as a function of light ellipticity, and reveal great similarities
between the response of the two systems to the external field. Both systems possess an angular shift in the momentum distributions, relative to
the predictions of the simple-man's model, and the magnitude of the shift is dependent
on light ellipticity. From analysis of the ATI spectra and electron density in the continuum, we obtain
further insight into the ionization dynamics and the ellipticity onset of the electron rescattering process.

\end{document}